\documentclass[9pt,twocolumn]{extarticle}
\usepackage{lastpage}
\usepackage{fancyhdr}
\fancypagestyle{plain}{%
  \fancyhf{}%
  \fancyfoot[C]{}%
  \fancyhead[HR]{\color{fresnelmiddleblue} \textbf{\thepage}}
  \fancyhead[HL]{\color{fresneldarkblue} \textbf{Institut Fresnel,} \color{fresnellightblue} \textbf{ATHENA Team}}
}

\usepackage[english]{babel}
\usepackage[scaled]{helvet}
\usepackage{authblk}
\usepackage{graphicx,xcolor}
\definecolor{color0}{RGB}{0,0,0}
\definecolor{red_orange}{RGB}{240,100,100}
\definecolor{guillgreen}{RGB}{50,240,50}
\definecolor{fredred}{RGB}{240,50,50}
\definecolor{fresneldarkblue}{RGB}{44,46,131} 
\definecolor{fresnelmiddleblue}{RGB}{34,124,192} 
\definecolor{fresnellightblue}{RGB}{0,165,226} 
\newcommand{\titlefont}{\color{fresnellightblue} \normalfont\sffamily\bfseries\fontsize{21}{23}\selectfont}

\usepackage{amsmath,amsfonts,amssymb}
\usepackage{booktabs,multirow,tabularx}
\usepackage[colorlinks=true, allcolors=blue]{hyperref}

\newcommand{\w}{\omega}

\newcommand{\Field}[2][]{\mathbf{#2}_{#1}}
\newcommand{\Fieldhat}[2][]{\mathbf{\hat{#2}}_{#1}}

\newcommand{\PD}[3][]{\frac{\partial^{#1} #2}{\partial #3^{#1}}}

\def\doubleunderline#1{\underline{\underline{#1}}}

\title{\titlefont Extracting an accurate model for permittivity from experimental data : Hunting complex poles from the real line}

\author[1,*]{M. Garcia-Vergara}
\author[]{G. Dem\'esy}
\author[]{F. Zolla}

\affil[]{Aix Marseille Universit\'e, CNRS, Centrale Marseille, Institut Fresnel UMR 7249, 13013 Marseille, France}

\affil[*]{Corresponding author: mauricio.garcia-vergara@fresnel.fr}
\date{\vspace{-5ex}}
\begin{document}
\maketitle
\begin{abstract}
In this letter, we describe a very general procedure
to obtain a causal fit of the permittivity of materials from experimental data with very few parameters.
Unlike other closed forms proposed in the literature, the particularity of this approach lies in its independence towards the material or frequency range at stake. Many illustrative numerical examples are given and the accuracy of the fitting is compared to other expressions in the literature.
\end{abstract}

\section{Introduction}

In this letter, we propose a general framework dedicated to the 
fitting of tabulated experimental data of complex permittivities 
\cite{palik1998handbook,weber2002handbook}.
When using time harmonic numerical methods in electromagnetism,
one sets a real frequency and uses the tabulated data, up to a simple 
interpolation, as it is.
However, in time domain methods (e.g. Finite difference \cite{taflove2000computational},
discontinuous Galerkin \cite{lu2004discontinuous}), 
the inverse Fourier Transform of the experimental data is needed since
frequency dispersion in the time domain is generally tackled through an 
extra differential equation involving the polarization vector. An 
analytical expression of the relative permittivity
$\hat{\epsilon}_r(\omega)$ fulfilling causality requirements has to be
extracted from the data given in frequency domain. Same considerations
hold when tackling generalized modal computations \cite{brule2016calculation}
of dispersive structures.
This problem is well known and several closed forms
have already been proposed: Drude and/or Drude-Lorentz model,
Debye model, critical points model \cite{etchegoin2006analytic}, or
a combination of these elementary resonances
\cite{barchiesi2015errata, deinega_effective_2012}. It is worth 
noticing that the form assumed is material dependent in the literature.
In this letter we do not assume any particular form for the permittivity. 
The only requirement is the causality principle handled via the general constitutive 
relation between the electric field $\mathbf{E}$ and the polarization vector
$\mathbf{P}_e$. Causality is also ensured by assuming numerically that the experimental data
satisfy appropriate parity requirements. We provide the 
details of the iterative least square approach, and provide numerical illustrations
for semi-conductors and metals.
\section{Mathematical formulation}
In a frequency dispersive material with negligible magnetization, the electric displacement $\Field[]{D}$ is not only influenced by the electric field $\Field[]{E}$, but also by the polarization vector $\Field[e]{P}$. A very general approach is to consider the following constitutive relation:
\begin{equation}
\label{eq:Const_relation}
	\sum_{l=0}^{N_{d}} q_{l} \PD[l]{\Field[e]{P}}{t} =\epsilon_0 \sum_{k=0}^{N_{n}} p_{k}\PD[k]{\Field[]{E}}{t}\,,
\end{equation}
where the reality of both $\Field[e]{P}$ and $\Field[]{E}$ requires $p_k$'s and $q_l$'s to be real numbers.   
Keeping causal solutions for $\Field[e]{P}$, it remains to carry out a Fourier transform with the following convention $\hat{f}(\omega)=\int_{\mathbb{R}} f(t) e^{-i \omega t} \, \mathrm{d}t$:
\begin{equation}
	\label{eq:FConst_relation}
	\left(\sum_{l=0}^{N_{d}} q_{l} (i\w)^l\right) \Fieldhat[e]{P}  =\epsilon_0 \left(\sum_{k=0}^{N_{n}} p_{k}(i\w)^k\right) \Fieldhat[]{E}\,.
\end{equation}
Then, the electric susceptibility $\hat{\chi}(\w)$ is given by
\begin{equation}
\label{eq:Rational_f_1}
	\hat{\chi}(\w) = \frac{\sum_{k=0}^{N_{n}} p_{k}(i\w)^k}{\sum_{l=0}^{N_{d}} q_{l} (i\w)^l}.
\end{equation}
Using the fact that the electric susceptibility is expressed as a rational function, it is convenient to divide all the coefficients by $q_0$, and calling $P_k = p_k/q_0$, $Q_l = q_l/q_0$ equation (\ref{eq:Rational_f_1}) takes the form
\begin{equation}
	\label{eq:Rational_f_2}
	\hat{\chi}(\w) = \frac{\sum_{k=0}^{N_{n}} P_{k}(i\w)^k}{\sum_{l=0}^{N_{d}} Q_{l} (i\w)^l}, \qquad Q_0 = 1.
\end{equation}
Consider now, some experimental data determined for instance by ellipsometry given by a set of corresponding points ($\w_m$, $\hat{\chi}^{Data}_m$) with $m = 0, \dots ,M-1$. For passive materials these data are such that $\omega_m \in \mathbb{R}^+$ and $\hat{\chi}_m^{Data} \in \mathbb{C}^-$ were $\mathbb{C}^- = \{\xi \in \mathbb{C} | \mathfrak{Im}\{\xi\}<0 \}$.
These data can be organized into the following vectors
\begin{equation}
\underline{\w}= (\w_0, \dots, \w_{M-1})^T, \qquad \underline{\hat{\chi}}^{Data} = (\hat{\chi}^{Data}_0, \dots, \hat{\chi}^{Data}_{M-1})^T\,.
\end{equation}
Let's suppose that each $\hat{\chi}^{Data}_m$ at $\w_m$ has a form as in Eq.~(\ref{eq:Rational_f_2}), that is:
\begin{equation}
	\label{eq:Rational_f_dis_1}
	\hat{\chi}^{Data}_m =  \frac{\sum_{k=0}^{N_{n}} P_{k}(i\w_m)^k}{\sum_{l=0}^{N_{d}} Q_{l} (i\w_m)^l}\,.
\end{equation}
After some elementary manipulations and remembering that $Q_0 = 1$, Eq.~(\ref{eq:Rational_f_dis_1}) reads:
\begin{equation}
	\label{eq:Rational_f_dis_2}
	\hat{\chi}^{Data}_m =  \sum_{n=0}^{N_{n}+N_{d}} r_{n} \xi_{m,n}\,,
\end{equation}
with
\begin{equation}
\label{eq:r_n}
r_n = \begin{cases}
     P_n      &\textrm{ if } n = 0,\dots,N_n\\
     Q_{n-N_n}&\textrm{ if } n = N_n+1,\dots,N_n+N_d
                         \end{cases}\,,
\end{equation}
and
\begin{equation}
\label{eq:xi_mn}
\xi_{m,n} = \begin{cases}
     		(i\w_m)^n                    &\textrm{ if } n = 0,\dots,N_n   \\
     		-\hat{\chi}^{Data}_m(i\w_m)^{n-N_n} &\textrm{ if } n = N_n+1,\dots,N_n+N_d
            \end{cases}\,.
\end{equation}
This can be rewritten in matrix form as: 
\begin{equation}
\underline{\hat{\chi}}^{Data} = \doubleunderline{\Xi} \textrm{  }\underline{r}\,,
\end{equation}
where $\underline{r}$ is the column vector with entries $r_n$, with $n = 0, \dots, N_n + N_d$ and $\doubleunderline{\Xi}$ is the $M\times (N_n+N_d+1)$ matrix with entries $\xi_{m,n}$. This overdetermined system can be solved in the sense of least squares \cite{strang_linear_2006}. That is, to find the vector $\underline{R}$ such that 
\begin{equation}
	\Vert  \doubleunderline{\Xi} \textrm{  }\underline{R} - \underline{\hat{\chi}}^{Data} \Vert_2 = \min_{\underline{r} \in \mathbf{R}^{N_n+N_d+1}} \Vert  \doubleunderline{\Xi} \textrm{  }\underline{r} - \underline{\hat{\chi}}^{Data} \Vert_2.
\end{equation}
This can be achieved, for instance, by the Householder transformation method \cite{skiba_metodos_2005}. In practice, we consider the entries of $\underline{r}$ to be complex, which relaxes our numerical scheme involving complex polynomials in ($i\omega$). The imaginary part of these numbers remains several orders of magnitude smaller than their real part.\\
Once the $p_k$'s and $q_l$'s coefficients have been determined, it is possible to obtain the poles and zeros of $\hat{\chi}(\w)$ by finding the roots of $\sum_{k=0}^{N_{d}} Q_{l}(i\w)^l$ and $\sum_{k=0}^{N_{n}} P_{k}(i\w)^k$ respectively. Let $\Omega_j$, $j = 1, \dots, N_d$ be the obtained poles 
then $\hat{\chi}(\w)$ can be expanded as follows \cite{petit_outil_1991}:
\begin{equation}
\label{eq:chi_PPg}
 \hat{\chi}(\w) = \sum_{j=1}^{N_d}\frac{A_j}{\w - \Omega_j} + g(\w)\,,
\end{equation}
where $g$ is an holomorphic function representing a non resonant term of $\hat{\chi}$ and is approximated by a polynomial of degree $N_n-N_d$. Assuming that this non resonant term is negligible, the amplitude coefficients $A_j$'s can be obtained via the Tetrachotomy method \cite{zolla_foundations_2005} or as in the case of this paper by another least squares procedure. Note that this latest assumption simply amounts to compelling the degree of the numerator to be smaller than the denominator's.
\section{Fitting data in practice}

\begin{figure}[htbp]
\centering
\fbox{\includegraphics[width=0.9\linewidth]{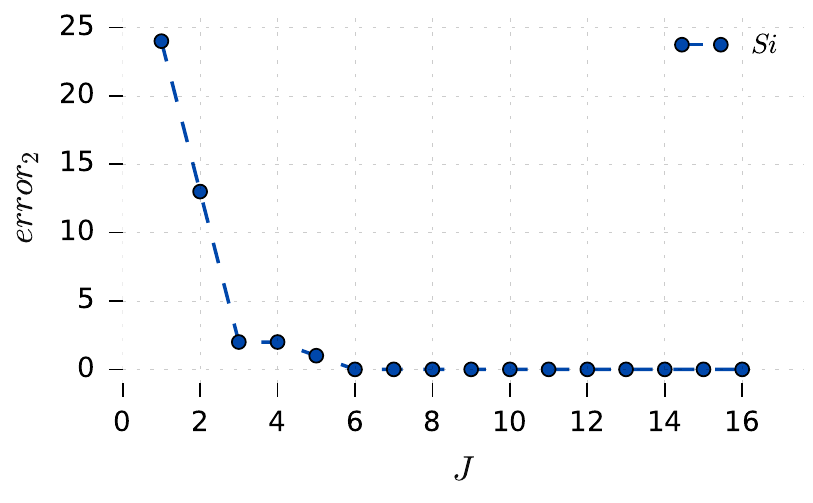}}
\caption{Norm two fitting error for Silicon as a function of the number of poles $J$}.
\label{fig:Si_error_2}
\end{figure}

\begin{figure}[htbp]
\centering
\fbox{\includegraphics[width=0.9\linewidth]{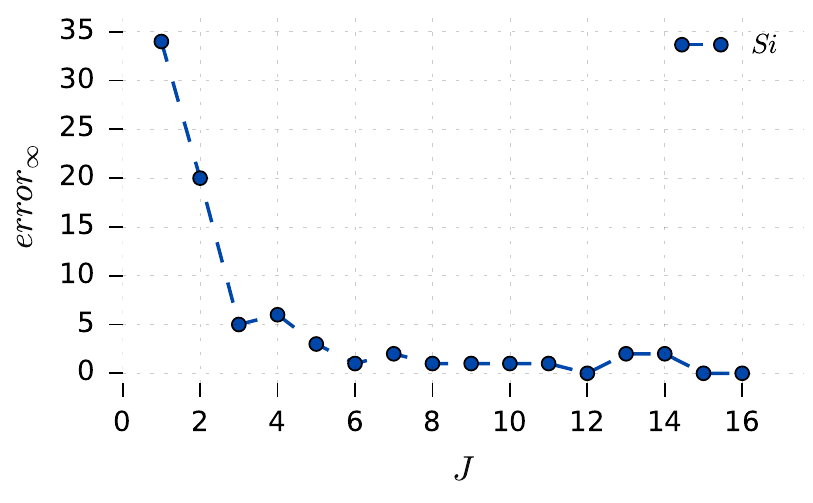}}
\caption{Norm infinity fitting error for Silicon as a function of the number of poles $J$}.
\label{fig:Si_error_inf}
\end{figure}

The first important step for fitting the data is to extend $\underline{\w}$ and $\underline{\hat{\chi}}^{Data}$, which are always given for positive frequencies only, so that the new vectors represent an electrical susceptibility with Hermitian symmetry. Second, we set $N_d = 2J$ for $J \in \mathbb{N}$ and $N_n \leq N_d$. This is done keeping in mind that each pole $\Omega_j$ have its corresponding symmetric $-\overline{\Omega}_j$ and in order to ensure that the non resonant function $g$ is at most a constant. In practice a good choice is to keep $N_n = N_d$, which is the approach we will follow in the sequel. Next, when the poles $\Omega_j$ and the associated amplitudes $A_j$ are computed, it is handy to sort these pairs by the modulus of $A_j$. Once the data sorted, the sign of the imaginary part of $\w_j$ has to be checked. In the case of our choice for the Fourier Transform, the imaginary part of physical poles $\w_j$ must be non negative. If the first $J_p$ pairs of poles $(\Omega_j, -\overline{\Omega_j})$, with $J_p \leq J$, have non negative imaginary part, then we can truncate the limit of the sum in Eq.~(\ref{eq:chi_PPg}), that is: 
\begin{equation}
	\hat{\chi}_{trunc}(\w) \approx \sum_{j=1}^{J_p}\frac{A_j}{\w - \Omega_j} - \frac{\overline{A}_j}{\w + \overline{\Omega}_j}.
\end{equation}
It is easy to see that this expression presents Hermitian symmetry. Finally, if the error between $\hat{\chi}_{trunc}(\underline{\w})$ and $\underline{\hat{\chi}}^{Data}$ according to a given norm is less than a certain tolerance, one can say the best fitting, according to this procedure, has been found. Otherwise it is necessary to repeat this procedure with $J+1$ points of poles and so on. It can be thought at first that the bigger the number of poles $J$ the better will be the fitting. This is not true in general. As an illustrative example, the two and infinity norms fitting errors for Si \cite{green_optical_1995} are shown in Figures \ref{fig:Si_error_2} and \ref{fig:Si_error_inf} respectively. Notice that while the norm two error decreases as $J$ increases, the norm infinity error does not exhibit a monotonic behavior. This is mainly due to the fact that the experimental data can exhibit measurement artefacts, for instance when switching from one source to another, and these small artefacts are revealed by the presence of spurious poles sometimes lying in the wrong (lower) half of the complex plane. These sharp spikes make the infinity norm increase once obvious poles are found.

\section{Results}
The following tables are present the values of the poles $\w_j$ and associated amplitudes $A_j = |A_j|\exp(i\phi_j)$ found when applying the method described above for different data sets. Also, the errors for $2-$norm and $\infty-$norm expressed in percentage are given. The materials considered in this paper are: Gold and Copper \cite{johnson_optical_1972}, Aluminum \cite{ordal_optical_1988}, Silver \cite{babar_optical_2015}, GaAs, GaP \cite{jellison_optical_1992} and Silicon \cite{green_optical_1995}. The values for Au and Cu can be read in Tables \ref{tab:Gold_poles} and \ref{tab:Copper_poles}, for wavelengths $\lambda$ in the range $\lambda:0.188-1.937 \mu m$. In the same way, Tables \ref{tab:Al_poles} and \ref{tab:Ag_poles} show the fitting values for Al and Ag with  $\lambda:0.667-200 \mu m$ and $\lambda:0.2066-12.40 \mu m$ respectively. Finally, for the case of semiconductors, GaAs and GaP parameters are given in Table~\ref{tab:GaAs_poles} and \ref{tab:GaP_poles} for $\lambda:0.234-0.840 \mu m$ while Si parameters are written in Table~\ref{tab:Si_poles} for $\lambda:0.25-1.0 \mu m$.These values were computed using the Python Code provided in (Ref.~\cite{mau_code_2016}).
\begin{table}[htbp]
\centering
\caption{\textbf{Au (Johnson and Christy)} $\boldsymbol{\lambda:0.188-1.937 \mu m}$}
\begin{tabular}{lll}
\hline
$\Omega_j$ [P $rad/s$]& $|A_j|$ & $\phi_j$ $[rad]$\\
\hline
$3.43E-01 + 5.21E-02i$ & $238.36$ & $-3.14$ \\
$4.56E+00 + 1.46E+00i$ & $9.83$ & $-2.12$ \\
$error_2$  (\%)& $3.01$  \\
$error_\infty$ (\%) & $1.27$  \\
\hline
\end{tabular}
  \label{tab:Gold_poles}
\end{table}
\begin{table}[htbp]
\centering
\caption{\textbf{Cu (Johnson and Christy)} $\boldsymbol{\lambda:0.188-1.937 \mu m}$}
\begin{tabular}{lll}
\hline
$\Omega_j$ [P $rad/s$]& $|A_j|$ & $\phi_j$ $[rad]$\\
\hline
$4.46E-01 + 4.61E-02i$ & $156.78$ & $-3.12$ \\
$3.11E+00 + 7.71E-01i$ & $5.16$ & $-1.07$ \\
$error_2$  (\%)& $6.70$  \\
$error_\infty$ (\%) & $2.88$  \\
\hline
\end{tabular}
  \label{tab:Copper_poles}
\end{table}
\begin{table}[htbp]
\centering
\caption{\textbf{Al (Ordal et al.)} $\boldsymbol{\lambda:0.667-200 \mu m}$ }
\begin{tabular}{lll}
\hline
$\Omega_j$ [P $rad/s$]& $|A_j|$ & $\phi_j$ $[rad]$\\
\hline
$3.24E-10 + 1.35E-03i$ & $4284.43$ & $-1.57$ \\
$1.13E-01 + 7.16E-02i$ & $228.84$ & $-3.08$ \\
$4.24E-01 + 7.94E-01i$ & $139.21$ & $0.69$ \\
$error_2$  (\%)& $8.36$  \\
$error_\infty$ (\%) & $11.55$  \\
\hline
\end{tabular}
  \label{tab:Al_poles}
\end{table}
\begin{table}[htbp]
\centering
\caption{\textbf{Ag (Babar et al)} $\boldsymbol{\lambda:0.2066-12.40 \mu m}$}
\begin{tabular}{lll}
\hline
$\Omega_j$ [P $rad/s$]& $|A_j|$ & $\phi_j$ $[rad]$\\
\hline
$-9.14E-16 + 6.52E-02i$ & $1818.56$ & $1.57$ \\
$8.37E+00 + 2.78E+00i$ & $6.83$ & $-2.44$ \\
$6.30E+00 + 4.72E-01i$ & $1.62$ & $-2.58$ \\
$6.73E+00 + 2.18E-01i$ & $0.39$ & $-1.55$ \\
$error_2$  (\%)& $1.71$  \\
$error_\infty$ (\%) & $1.87$  \\
\hline
\end{tabular}
  \label{tab:Ag_poles}
\end{table}
\begin{table}[htbp]
\centering
\caption{\textbf{GaAs (Jellison et al.)} $\boldsymbol{\lambda:0.234-0.840 \mu m}$}
\begin{tabular}{lll}
\hline
$\Omega_j$ [P $rad/s$]& $|A_j|$ & $\phi_j$ $[rad]$\\
\hline
$7.17E+00 + 8.55E-01i$ & $18.54$ & $-2.92$ \\
$4.65E+00 + 1.00E+00i$ & $12.34$ & $-2.96$ \\
$4.30E+00 + 2.57E-01i$ & $2.37$ & $-1.54$ \\
$7.66E+00 + 2.40E-01i$ & $1.79$ & $1.29$ \\
$error_2$  (\%)& $3.13$  \\
$error_\infty$ (\%) & $6.23$  \\
\hline
\end{tabular}
  \label{tab:GaAs_poles}
\end{table}
\begin{table}[htbp]
\centering
\caption{\textbf{GaP (Jellison et al.)} $\boldsymbol{\lambda:0.234-0.840 \mu m}$}
\begin{tabular}{lll}
\hline
$\Omega_j$ [P $rad/s$]& $|A_j|$ & $\phi_j$ $[rad]$\\
\hline
$7.60E+00 + 7.60E-01i$ & $20.56$ & $-3.10$ \\
$5.64E+00 + 2.40E-01i$ & $5.02$ & $-2.87$ \\
$6.19E+00 + 3.51E-01i$ & $1.96$ & $-2.62$ \\
$4.32E+00 + 4.49E-01i$ & $0.68$ & $-1.87$ \\
$error_2$  (\%)& $3.16$  \\
$error_\infty$ (\%) & $6.78$  \\

\hline
\end{tabular}
  \label{tab:GaP_poles}
\end{table}
\begin{table}[htbp]
\centering
\caption{\textbf{Si (Green and Keevers)} $\boldsymbol{\lambda:0.25-1.0 \mu m}$}
\begin{tabular}{lll}
\hline
$\Omega_j$ [P $rad/s$]& $|A_j|$ & $\phi_j$ $[rad]$\\
\hline
$7.99E+00 + 1.83E+00i$ & $13.57$ & $2.98$ \\
$6.54E+00 + 3.74E-01i$ & $11.50$ & $2.78$ \\
$5.49E+00 + 6.65E-01i$ & $10.51$ & $-2.46$ \\
$5.12E+00 + 1.68E-01i$ & $4.02$ & $-2.46$ \\
$error_2$  (\%)& $1.08$  \\
$error_\infty$ (\%) & $3.08$  \\
\hline
\end{tabular}
  \label{tab:Si_poles}
\end{table}
\section{Validation}
In order to test the validity of our approach, two comparisons have been realized between our results and the ones found in the literature. In the case of metals, the results reported by Barchiesi and Grosges (B\&G) in \cite{barchiesi2015errata} for Gold (using experimental data from \cite{johnson_optical_1972}). In this case we set $J$ = 8 and truncate the sum up to $J_p = 2$ (the same number of poles considered by B\&G). The errors using $2-$norm and $\infty-$norm expressed in percentage obtained by us are 3.01\% and 1.27\%, while for B\&G are 9.98\% and 6.65\% respectively. For semiconductors, the Deinega and John (D\&J) fitting parameters obtained for Silicon by considering two poles allow to compute a $2-$norm error of 8.5\% and a $\infty-$norm error of 15.63\%. On the other hand, our approach setting $J = 6$ and $J_p = 4$ (the double of poles than D\&J) allow us to compute corresponding two and infinity norm errors of 1.08\% and 3.08\%.
\begin{figure}[htbp]
\centering
\fbox{\includegraphics[width=0.9\linewidth]{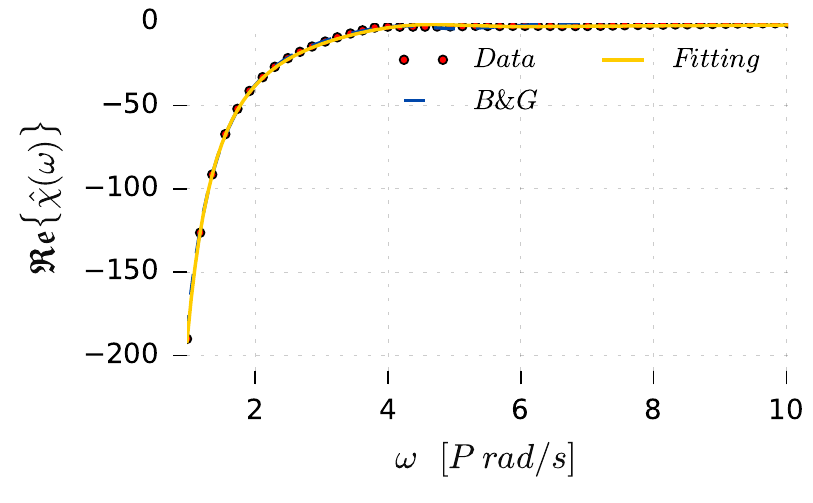}}
\caption{Comparison between the experimental data, the approach by Barchiesi and Grosges (B\&G) and our fitting for the real part of $\hat{\chi}$ for Gold.}
\label{fig:Au_re}
\end{figure}
\begin{figure}[htbp]
\centering
\fbox{\includegraphics[width=0.9\linewidth]{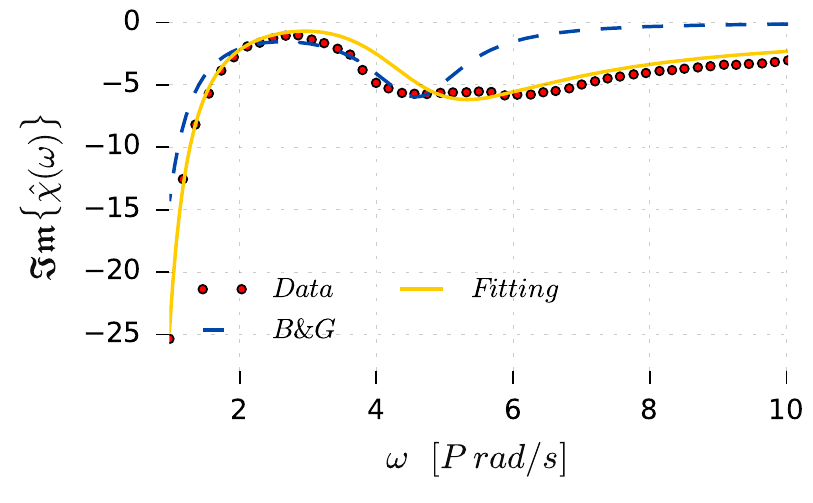}}
\caption{Comparison between the experimental data, the approach by Barchiesi and Grosges (B\&G) and our fitting for the imaginary part of $\hat{\chi}$ for Gold.}
\label{fig:Au_im}
\end{figure}
\begin{figure}[htbp]
\centering
\fbox{\includegraphics[width=0.9\linewidth]{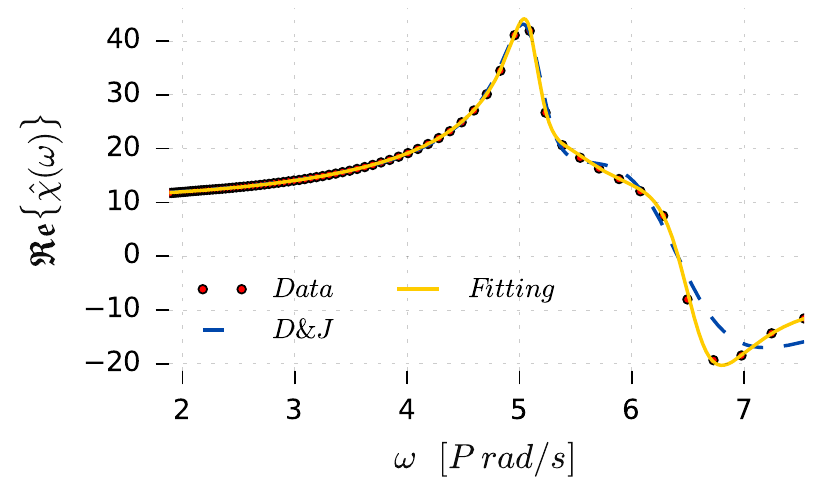}}
\caption{Comparison between the experimental data, the approach by Daneiga and John (D\&J) and our fitting for the real part of $\hat{\chi}$ for Silicon.}
\label{fig:Si_re}
\end{figure}
\begin{figure}[htbp]
\centering
\fbox{\includegraphics[width=0.9\linewidth]{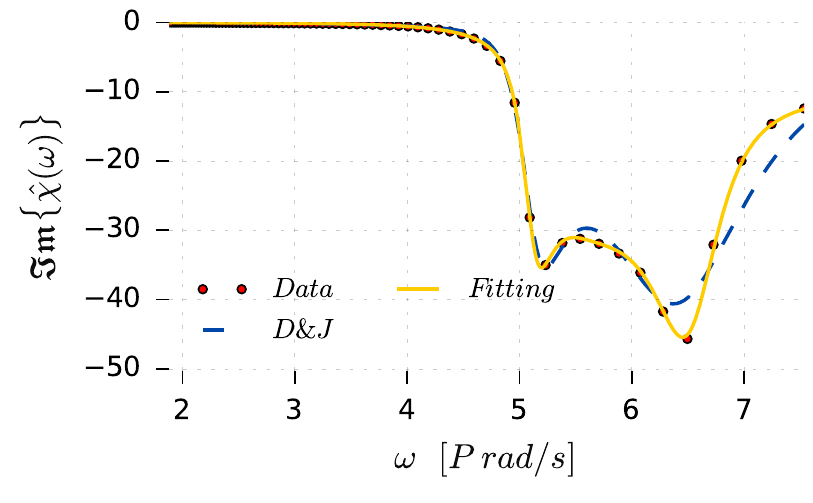}}
\caption{Comparison between the experimental data, the approach by Daneiga and John (D\&J) and our fitting for the imaginary part of $\hat{\chi}$ for Silicon.}
\label{fig:Si_im}
\end{figure}
\section{Conclusions}
In this letter, we proposed a simple yet systematic procedure for fitting experimental data of permittivities of resonant materials such as metals and semiconductors in the visible range. This procedure does not assume \textit{a priori} any particular shape for the electric susceptibility $\hat{\chi}$. The final expression obtained for the permittivity preserves causality and stability. This fitting is more accurate than those presented in the reviewed literature. It can be used as it is in numerical codes such as FDTD.
\bibliographystyle{ieeetr}
\bibliography{biblio}

\begin{thebibliography}{10}

\bibitem{palik1998handbook}
E.~D. Palik, {\em Handbook of optical constants of solids}, vol.~3.
\newblock Academic press, 1998.

\bibitem{weber2002handbook}
M.~J. Weber, {\em Handbook of optical materials}, vol.~19.
\newblock CRC press, 2002.

\bibitem{taflove2000computational}
A.~Taflove and S.~C. Hagness, {\em Computational electrodynamics}.
\newblock Artech house publishers, 2000.

\bibitem{lu2004discontinuous}
T.~Lu, P.~Zhang, and W.~Cai, ``Discontinuous galerkin methods for dispersive
  and lossy maxwell's equations and pml boundary conditions,'' {\em Journal of
  Computational Physics}, vol.~200, no.~2, pp.~549--580, 2004.

\bibitem{brule2016calculation}
Y.~Br{\^u}l{\'e}, B.~Gralak, and G.~Dem{\'e}sy, ``Calculation and analysis of
  the complex band structure of dispersive and dissipative two-dimensional
  photonic crystals,'' {\em JOSA B}, vol.~33, no.~4, pp.~691--702, 2016.

\bibitem{etchegoin2006analytic}
P.~G. Etchegoin, E.~Le~Ru, and M.~Meyer, ``An analytic model for the optical
  properties of gold,'' {\em The Journal of chemical physics}, vol.~125,
  no.~16, p.~164705, 2006.

\bibitem{barchiesi2015errata}
D.~Barchiesi and T.~Grosges, ``Errata: Fitting the optical constants of gold,
  silver, chromium, titanium and aluminum in the visible bandwidth,'' {\em
  Journal of Nanophotonics}, vol.~8, no.~1, p.~089996, 2015.

\bibitem{deinega_effective_2012}
A.~Deinega and S.~John, ``Effective optical response of silicon to sunlight in
  the finite-difference time-domain method,'' {\em Optics Letters}, vol.~37,
  pp.~112--114, Jan. 2012.

\bibitem{strang_linear_2006}
G.~Strang, {\em Linear {Algebra} and {Its} {Applications}}.
\newblock Thomson, Brooks/Cole, 2006.
\newblock Google-Books-ID: 8QVdcRJyL2oC.

\bibitem{skiba_metodos_2005}
Y.~Skiba, {\em Metodos {Y} {Esquemas} {Numericos} : {Un} {Analisis}
  {Computacional}}.
\newblock UNAM, June 2005.

\bibitem{petit_outil_1991}
R.~Petit, {\em L'outil mathématique: distributions, convolution,
  transformations de {Fourier} et de {Laplace}, fonctions d'une variable
  complexe, fonctions eulériennes}.
\newblock Masson, 1991.

\bibitem{zolla_foundations_2005}
F.~Zolla, {\em Foundations of {Photonic} {Crystal} {Fibres}}.
\newblock Imperial College Press, Jan. 2005.
\newblock Google-Books-ID: iVZXwXDswv0C.

\bibitem{green_optical_1995}
M.~A. Green and M.~J. Keevers, ``Optical properties of intrinsic silicon at 300
  {K},'' {\em Progress in Photovoltaics: Research and Applications}, vol.~3,
  pp.~189--192, Jan. 1995.

\bibitem{johnson_optical_1972}
P.~B. Johnson and R.~W. Christy, ``Optical {Constants} of the {Noble}
  {Metals},'' {\em Physical Review B}, vol.~6, pp.~4370--4379, Dec. 1972.

\bibitem{ordal_optical_1988}
M.~A. Ordal, R.~J. Bell, R.~W. Alexander, L.~A. Newquist, and M.~R. Querry,
  ``Optical properties of {Al}, {Fe}, {Ti}, {Ta}, {W}, and {Mo} at
  submillimeter wavelengths,'' {\em Applied Optics}, vol.~27, pp.~1203--1209,
  Mar. 1988.

\bibitem{babar_optical_2015}
S.~Babar and J.~H. Weaver, ``Optical constants of {Cu}, {Ag}, and {Au}
  revisited,'' {\em Applied Optics}, vol.~54, pp.~477--481, Jan. 2015.

\bibitem{jellison_optical_1992}
G.~E. Jellison, ``Optical functions of silicon determined by two-channel
  polarization modulation ellipsometry,'' {\em Optical Materials}, vol.~1,
  pp.~41--47, Jan. 1992.

\bibitem{mau_code_2016}
M.~Garcia-Vergara, G.~Dem{\'e}sy, and F.~Zolla, ``Hunting complex poles,''
  (GitHub 2016) [retrieved 6 Dec 2016].
\newblock https://git.io/v14Jb.

\end{thebibliography}
\end{document}